\documentclass[prx,twocolumn,showpacs,superscriptaddress,preprintnumbers,amssymb]{revtex4-2} 
\usepackage{graphicx}
\usepackage{latexsym}
\usepackage{amssymb}
\usepackage{amsmath}
\usepackage{amsfonts} 
\usepackage{upgreek}
\usepackage{bm}
\usepackage{multirow}
\usepackage{enumitem}
\usepackage{color}
\usepackage[colorlinks, citecolor=blue]{hyperref}
\usepackage{physics}
\usepackage{tcolorbox}
\usepackage{manfnt}
\usepackage{lipsum}
\usepackage[ruled,linesnumbered]{algorithm2e}
\usepackage{tabularx}
\usepackage{nicematrix}
\NiceMatrixOptions{
code-for-first-row = \color{blue} ,
code-for-last-row = \color{blue} ,
code-for-first-col = \color{blue} ,
code-for-last-col = \color{blue}
}

\usepackage{amsthm}
\theoremstyle{definition}

\newcommand{\cO}{ {\cal O} }
\newcommand{\cD}{ {\cal D} }
\newcommand{\cE}{ {\cal E} }
\newcommand{\cG}{ {\cal G} }

\newcommand{\be}{\mathbf{e}}

\newcommand{\tc}{{\textrm{tc}}}

\newcommand{\opz}{\overline{p^z_{\bm{m} \vphantom{\tilde{\bmm}}}}}
\newcommand{\opx}{\overline{p^x_{\tilde{\bm{m}}}}}
\newcommand{\opi}{\overline{p^i_{\bmm}}}

\newcommand{\bap}{{\bm{a'}}}
\newcommand{\ba}{{\bm{a}}}
\newcommand{\balp}{{\bm{\alpha}}}
\newcommand{\balpp}{{\bm{\alpha'}}}
\newcommand{\bb}{{\bm{b}}}

\newcommand{\bsigma}{{\bm{\sigma}}}

\newcommand{\bs}{{\bm{s}}}

\newcommand{\tbm}{{\tilde{\bm{m}}}}
\newcommand{\bmm}{{\bm{m}}}

\newcommand{\bl}{{\bm{l}}}

\newcommand{\bmeta}{{\bm{\eta}}}

\newcommand{\eqnref}[1]{Eq.\,\eqref{#1}}

\newcommand{\figref}[1]{Fig.\,\ref{#1}}

\newcommand{\rd}{\partial}

\newcommand{\vdagger}{{\vphantom{\dagger}}}

\newcommand{\supmat}{\text{Supplementary Material}}

\renewcommand{\tr}{\mathrm{tr}}

\newcounter{protocol}

\begin{document}

\title{Exact Calculations of Coherent Information for Toric Codes under Decoherence: Identifying the Fundamental Error Threshold}
\author{Jong Yeon Lee}\email{jongyeon@illinois.edu}
\affiliation{Department of Physics, University of California, Berkeley, California 94720, USA}
\affiliation{Department of Physics and Institute of Condensed Matter Theory, University of Illinois at Urbana-Champaign, Urbana, Illinois 61801, USA}

\date{\today}
\begin{abstract}
The toric code is a canonical example of a topological error-correcting code. Two logical qubits stored within the toric code are robust against local decoherence, ensuring that these qubits can be faithfully retrieved as long as the error rate remains below a certain threshold. Recent studies have explored such a threshold behavior as an intrinsic information-theoretic transition, independent of the decoding protocol. These studies have shown that information-theoretic metrics, calculated using the Renyi (replica) approximation, demonstrate sharp transitions at a specific error rate. However, an exact analytic expression that avoids using the replica trick has not been shown, and the connection between the transition in information-theoretic capacity and the random bond Ising model (RBIM) has only been indirectly established. In this work, we present the first analytic expression for the coherent information of a decohered toric code, thereby establishing a rigorous connection between the fundamental error threshold and the criticality of the RBIM.
\end{abstract}

\maketitle

\emph{Introduction.}--- Protecting data from errors is paramount in information transmission and utilization~\cite{Shannon}. This requirement poses a serious challenge for quantum information, which is inherently fragile and non-clonable~\cite{noclone}. 
Consequently, the design and characterization of robust quantum memories and processors has become a central objective in both applied and theoretical research~\cite{Shor1995, Knill, Kitaev_1997, Aharonov1997, Gottesman1998}.

In particular, the error tolerance of the toric code—a prototypical topological quantum error-correcting code—has been extensively investigated~\cite{SurfaceCode1998, Dennis2002TQM}.
Dennis et al.~\cite{Dennis2002TQM} argued that the error threshold of the maximum entropy decoder under Pauli errors maps to the critical temperature of a random bond Ising model (RBIM) along the Nishimori line~\cite{Nishimori, Nishimori2}.
Following this approach, decoding problems of various noisy quantum codes have been associated with statistical models and their transitions~\cite{Katzgraber2009, PhysRevLett.104.050504, Bombin2012, kovalev2014spin, Kubica2018, Chubb2021}.
However, the decoding threshold depends on the chosen algorithm, so the resulting threshold may differ from the fundamental limit.

To address the fundamental threshold, recent works analyzed the toric code under Pauli errors without focusing on any particular decoder~\cite{Fan2023, Lee2023PRXQ}. They identified critical points in information-theoretic measures, using the replica method to set an upper bound on the error threshold.
They also suggested that extrapolating these critical error rates to the $n\,{\rightarrow}\,1$ limit aligns with the RBIM’s Nishimori-line critical point.
Similar transition behavior induced by decoherence has been studied 
in different settings~\cite{LeeNishimori2022, PhysRevLett.131.200201, LeeSPT2022, Chen:2023tfg, sang2023mixedstate, colmenarez2024accurate} or 
via different approaches, such as separability criterion~\cite{HastingsSeparability2011, chen2023separability}. 
However, the exact behavior of quantum information in a decohered toric code remains unclear.

In this work, we derive the toric code’s fundamental error threshold under Pauli errors without approximations, establishing a rigorous connection to the random-bond Ising model (RBIM).
Our approach centers on the analytic calculation of the coherent information, which quantifies the amount of decodable quantum information. Since robust coherent information gives both a necessary and sufficient condition for error correction~\cite{coherentInfo1,coherentInfo2}, its transition point sets a fundamental limit on the code’s decoding capacity. This differs from earlier work on maximum-likelihood or entropy decoders~\cite{Katzgraber2009, PhysRevLett.104.050504, Bombin2012, kovalev2014spin, Kubica2018, Chubb2021}, which used the free energy of the corresponding statistical mechanics model as a threshold criterion. We show that this free energy benchmark is only a rough estimate; its divergence does not ensure perfect decoding. Instead, coherent information emerges as the more precise and reliable metric.

\begin{figure}[!t]
    \includegraphics[width=1\columnwidth]{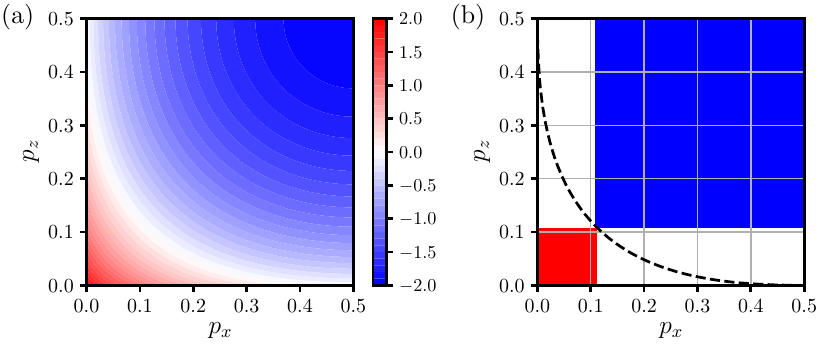}
    \caption{{\bf Coherent Information under Pauli-$Z$ and $X$ errors} for {\bf (a)} Two raw physical qubits $I_c^\textrm{raw}$ and {\bf (b)} Two logical qubits of the toric code state $I_c^\tc$ in the thermodynamic limit. 
    The dashed line in (b) is the contour of $I_c^\textrm{raw} = 0$, which passes the point $(0.1100, 0.1100)$. This point is very close to the critical point of the Nishimori line $(0.1094,0.1094)$~\cite{NishimoriPoint}.  }
    \label{fig:2}
\end{figure}

\begin{figure}[!t]
    \includegraphics[width=1\columnwidth]{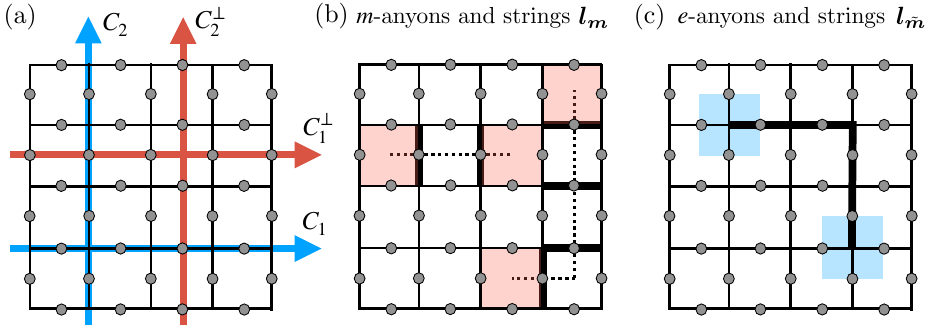}
    \caption{ {\bf Conventions.} {\bf(a)} Cycles along the edges of the original (blue) and dual (red) lattices $C_i$ and $C_i^\perp$ respectively. {\bf(b)} A set of red plaquettes $\bmm$ in the original lattice to denote $m$-anyons and corresponding string $\bl^\bmm$ of Pauli-$Z$s. {\bf(c)} A set of blue plaquettes $\tilde{\bmm}$ in the dual lattice to denote $e$-anyons and corresponding string (thick black lines) $\bl^\tbm$ of Pauli-$X$s.}
    \label{fig:1}
\end{figure}

\emph{Model.}--- We consider the toric code on torus, whose Hamiltonian is defined as
\begin{align} \label{eq:toric}
        H &= - \sum_v A_v - \sum_p B_p
\end{align}
where $A_v := \prod_{e \ni v } {Z}_e$ and $B_p :=\prod_{e \in p} X_e$. 
Its ground state satisfies $A_v = B_p = 1$ and is four-fold degenerate on a torus, encoding two logical qubits. Pauli operators acting on the logical space are $\overline{\bm{X}}_i\,{:=}\,\prod_{e \in C_i} X_e$ and $\overline{\bm{Z}}_i\,{:=}\,\prod_{e \in C^\perp_{\bar{i}}} Z_e$ where $C_i^{\vphantom{\perp}}$ ($C_i^\perp$) is a (dual) cycle along the $i$-th axis. Here, $C^\perp_{\bar{1}}\,{=}\,C^\perp_2$ and $C^\perp_{\bar{2}}\,{=}\,C^\perp_1$ such that $\overline{\bm{Z}}_i \overline{\bm{X}}_i\,{=}\,{-}\overline{\bm{X}}_i \overline{\bm{Z}}_i $. While $C_i$ is defined along the edges, $C_i^\perp$ is along the dual edges as in \figref{fig:1}(a).

Next, we maximally entangle the toric code’s two logical qubits with two reference qubits.
To enforce Bell-type maximal entanglement, we set $Z^\textrm{r}_i  \overline{\bm{Z}}_i^{\vphantom{r}} = X^\textrm{r}_i  \overline{\bm{X}}_i^{\vphantom{r}} = 1$ for $i = 1,2$, where $Z^\textrm{r}_{i}$ and $X^\textrm{r}_{i}$ act on reference qubits.
Label the toric code system as $Q$ and the reference as $R$. The full density matrix is
\begin{align} \label{eq:dm_definition}
\rho_{RQ} &:= 4 \hspace{-5pt} \prod_{i \in\{1,2\}} \hspace{-3pt} \qty(\frac{1 + Z_i \overline{\bm{Z}}_i}{2}) \qty(\frac{1 + X_i \overline{\bm{X}}_i}{2}) (\mathbb{I}_R \otimes \rho_Q) \nonumber \\ \rho_Q &:= \tr_R(\rho_{RQ}) = \frac{1}{4} \prod_v \qty(\frac{1+A_v}{2}) \prod_p \qty(\frac{1+B_p}{2})
\end{align}
The coherent information of the system $Q$ under a decoherence channel $\cE$ is defined as~\cite{coherentInfo1, coherentInfo2} 
\begin{align} \label{eq:coherentInfo}
    I_c(R,Q; \cE) &:= S( \cE[\rho_{Q}] ) - S( (\textrm{id}_R \otimes \cE)[\rho_{RQ}]).
\end{align}
By maximally entangling the logical qubits with $R$, we quantify how much quantum information remains recoverable in $Q$. 
By the quantum data-processing inequality~\cite{coherentInfo1}, coherent information never increases under a quantum channel. 
Hence, if a decoder $\cD$ successfully recovers the logical qubits, it also restores entanglement with $R$, so $I_c(R,Q;\cD\circ \cE)=2 \log 2$.
The data-processing inequality then implies this can occur only if $I_c$ is unchanged by $\cE$. 
Thus, the coherent information under $\cE$ provides a fundamental bound on recoverable information, independent of the decoding protocol.

Throughout the paper, we consider decoherence channels for  uncorrelated Pauli-$X$ and $Z$ errors (see \supmat for correlated case):
\begin{align} \label{eq:dephasing}
    \cE^z_e: \rho &\rightarrow (1-p_z)\rho + p_z Z^\vdagger_e \rho Z^\dagger_e, \quad \cE^z = \prod_e \cE^z_e \nonumber \\
    \cE^x_e: \rho &\rightarrow (1-p_x)\rho + p_x X^\vdagger_e \rho X^\dagger_e, \quad \cE^x = \prod_e \cE^x_e.
\end{align}
We aim to diagonalize $\cE[\rho_Q]$ and $(\mathrm{id}\otimes \cE)[\rho_{RQ}]$ to compute \eqnref{eq:coherentInfo}. 
A convenient expansion is
\begin{align} \label{eq:error_expression}
    \cE[\rho] = \sum_{\bl_x, \bl_z}  \prod_{i\in x,z}   (1-p_i)_{\vphantom{i}}^{2N - |\bl_i|}  p_i^{|\bl_i|} \cdot  X^\vdagger_{\bl_x} Z^\vdagger_{\bl_z} \rho Z^\dagger_{\bl_z} X^\dagger_{\bl_x}
\end{align}
where $\bl_i\,{=}\,{l_{i,e}}$ with $l_{i,e}\,{\in}\,{0,1}$ marks the edges in the string, $|\bl_i|\,{=}\,\sum_e l_{i,e}$, $X_\bl\,{:=}\,\prod_{e} X^{l_e}_e$ and $Z_\bl\,{:=}\,\prod_{e} Z^{l_e}_e$, and $N$ is the number of vertices.

\emph{Diagonalization}.--- Consider the toric code state $|\psi^\tc_0\rangle$ for which $\overline{\bm{X}}_{1,2}\,{=}\,1$ and define $\rho_0\,{=}\,|\psi_0 \rangle \langle \psi_0 |$. The key observation is that $\cE[\rho_0]$ commutes with every local stabilizer $A_v$ and $B_p$.
Thus, we can diagonalize $\cE[\rho_0]$ in the stabilizer eigenbasis. Because there are $2(N\,{-}\,1)$ independent stabilizers for $2N$ qubits, $2^{2N}$ dimensional Hilbert space splits into $2^{2N-1}$ stabilizer sectors, each of which carries a $2^2$ dimensional logical subspace. 
Using the eigenbasis of $\overline{\bm{X}}_{1,2}$ for this subspace, non-trivial elements of a decohered density matrix can be referred to by the following labels:
\begin{itemize}[leftmargin=13pt]
    \item Eigenvalues of $B_p$s $\bmm\,{=}\,\{ m_p \}$. A Pauli-$Z$ string $\bl^\bmm$ on the dual-lattice edges creates the $m$-anyon pattern $\bmm$, obeying $\rd_{\scriptscriptstyle{\perp}} \bl^\bmm \,{\equiv}\,\bmm$~\footnote{The boundary operator $\rd$ ($\rd^{\scriptscriptstyle{\perp}}$) maps a set of values defined on (dual) edges into a set of values defined on (dual) vertices; $(\rd \bl)_v\,{:=}\,\sum_{v' \ni v} l_{(v,v')}$. Note that the plaquettes of the original lattice correspond to the vertices of the dual lattice.
    }. See \figref{fig:1}(b). 
    \item Eigenvalues of $A_v$s $\tilde{\bmm}\,{=}\,\{ m_v \}$. A Pauli-$X$ string  $\bl_\tbm$ on the original lattice creates the $e$-anyon configuration $\tbm$, obeying $\rd \bl^\tbm\,{\equiv}\,\tbm$. See \figref{fig:1}(c). 
    \item Eigenvalues of $(\overline{\bm{X}}_1,\overline{\bm{X}}_2)$ labels each logical subspace. We parameterize them via $\ba\,{=}\,(a_1,a_2)$ and $\bap\,{=}\,(a'_1,a'_2)$ such that $\bm{X}_i\,{=}\,e^{\mathrm{i} \pi a_i}$.
\end{itemize}
Hence, we expand $\cE[\rho_0]$ in the following basis: 
\begin{align} \label{eq:basis_expansion}
    \rho_{\bmm \tilde{\bmm}}^{\ba, \bap} &:= X_{\bl^\tbm}^\vdagger 
 Z_{\bl^{\bmm}}^\vdagger \overline{\bm{Z}}_{\ba}^\vdagger   |\psi_0 \rangle \langle \psi_0 |\overline{\bm{Z}}_{\bap}^\dagger  Z_{\bl^{\bmm}}^\dagger  X_{\bl^\tbm}^\dagger, 
\end{align}
where $\overline{\bm{Z}}_{\ba}\,{:=}\,\overline{\bm{Z}}_1^{a_1}\overline{\bm{Z}}_2^{a_2}$. 
Since $\rho_{\bmm \tilde{\bmm}}^{\ba, \bap}$s are orthonormal, 
\begin{align}
 \cE[\rho_0] &= \sum_{\bmm \tbm \ba \bap}  \tr( \cE[\rho_0] (\rho_{\bmm \tilde{\bmm}}^{\ba, \bap})^\dagger ) \cdot \rho_{\bmm \tilde{\bmm}}^{\ba, \bap}.
\end{align}
Each coefficient is found by plugging in \eqnref{eq:error_expression}:
\begin{align} \label{eq:matrixel}
    &\tr( \cE[\rho_0] (\rho_{\bmm \tilde{\bmm}}^{\ba, \bap})^\dagger )  = \sum_{\bl_x, \bl_z} \Big(\prod_{i\in x,z}   (1-p_i)_{\vphantom{i}}^{2N - |\bl_i|}  p_i^{|\bl_i|} \Big) \nonumber \\
    &   \qquad \qquad \quad \times \abs{\langle \psi_0 | Z_{\bl_z}^\dagger X_{\bl_x}^\dagger X_{\bl^\tbm}^\vdagger 
 Z_{\bl^{\bmm}}^\vdagger \overline{\bm{Z}}_{\ba}    | \psi_0 \rangle  }^2 \delta_{\ba, \bap}.
\end{align}
Nonzero overlaps occur only when:
$(i)$ $\rd_{\scriptscriptstyle{\perp}} (\bl^\bmm\,{-}\,\bl_z)\,{\equiv}\,\bm{0}$ and  $\rd(\bl^\tbm\,{-}\,\bl_x)\,{\equiv}\,\bm{0}$, and 
$(ii)$ $\bl_z\,{-}\,\bl^\bmm$ is homologically equivalent to the loop $\bm{L}^\perp_\ba$ defined by the vector $\ba$. If $a_i\,{\neq}\,0$, $\bm{L}^\perp_\ba$ has a non-trivial dual cycle along $i$-th direction. 
Symbolically, this is represented as $[\bl_z\,{-}\,\bl^\bmm]\,{\equiv}\,[\bm{L}^\perp_\ba]\,{=}\,\ba$ where the bracket notation indicates the homotopy class. Thus, we get
\begin{align} \label{eq:decomposition}
    \tr( \cE[\rho_0] (\rho_{\bmm \tilde{\bmm}}^{\ba, \bap})^\dagger )  &= \hspace{-5pt} \sum_{ \substack{ \rd_\perp (\bl - \bl^\bmm) = 0, \\
     [\bl - \bl^\bmm]=\ba} }  \hspace{-7pt}  (1-p_z)^{2N - |\bl|} p_z^{|\bl|}   \,\, \delta_{\ba,\bap} \nonumber \\
    & \times \hspace{-5pt} \sum_{ \substack{ \rd(\bl' - \bl^\tbm)=0}} \hspace{-7pt}  (1-p_x)^{2N - |\bl'|} p_x^{|\bl'|}.
\end{align}

Next, we parameterize $\bl$. Because $2^{N+1}$ constraints act on $2^{2N}$ possible configurations, exactly $2^{N-1}$ solutions for $\bl$ remain.
One particular solution is $\bl\,{\equiv}\,\bl^\bmm\,{+}\,\bm{L}^\perp_\ba$. 
Defining $\bs^{\bmm,\ba}\,{:=}\,\exp(i \pi (\bl^\bmm\,{+}\,\bm{L}^\perp_\ba))$, all $\bl$ satisfying $\rd_{\scriptscriptstyle{\perp}} (\bl_z\,{-}\,\bl^\bmm)\,{=}\,0$ and $[\bl_z\,{-}\,\bl^\bmm]\,{\equiv}\,\ba$ can be parametrized by $(-1)^{l_e}\,{=}\,s^{\bmm,\ba}_e \sigma_v \sigma_{v'}$, where $e\,{=}\,(v,v')$ and each $\sigma_v \in \{\pm\}$ lives on a vertex. Although there are $2^N$ configurations of $\bsigma\,{=}\,\{\sigma_v\}$, $\bsigma$ and $-\bsigma$ yield the same $\bl$, giving precisely $2^{N-1}$ distinct configurations. 
Noting $\sum_e (-1)^{l_e} \,{=}\,2(N\,{-}\,|\bl|)$, the first factor in \eqnref{eq:decomposition} becomes:
\begin{align} \label{eq:pz}
    &\hspace{-5pt}\sum_{ \substack{ \rd_{\scriptscriptstyle{\perp}}(\bl - \bl^\bmm) = 0, \\
     [\bl - \bl^\bmm]=\ba} }  \hspace{-13pt}  (1-p_z)^{2N - |\bl|} p_z^{|\bl|} = p_z^N (1-p_z)^N  \hspace{-15pt}  \sum_{ \substack{ \rd_{\scriptscriptstyle{\perp}}(\bl - \bl^\bmm) = 0, \\
     [\bl - \bl^\bmm]=\ba} }  \hspace{-13pt} e^{\beta_z \sum_e (-1)^{l_e} } \nonumber \\
    &\qquad =\frac{\sum_{\bsigma} e^{\beta_z \sum_e s^{\bmm,\ba}_{e} \sigma_v \sigma_{v'} }}{2(2\cosh \beta_z)^{2N}}  =\frac{ Z[\bs^{\bmm,\ba}, \beta_z] }{2(2\cosh \beta_z)^{2N}} 
\end{align}
where $e^{-2\beta_z}\,{=}\,p_i/(1\,{-}\,p_i)$, and $Z[\bs^{\bmm,\ba},\beta_z]$ is the RBIM partition function with bond configuration $\bs^{\bmm,\ba}$ at inverse temperature $\beta_z$~\cite{SM}.

The second factor in \eqnref{eq:decomposition} can be analyzed similarly, but there is subtlety. First, the parametrization of $\bl'$ satisfying $\rd(\bl^\tbm\,{-}\,\bl')\,{=}\,0$ (and the corresponding Ising model) lives on the dual lattice. 
Moreover, since no condition is imposed on the homotopy class of $\bl'-\bl^\tbm$, we write $\bl'$ in terms of both $\tilde{\bsigma}\,{=}\,\{\tilde{\sigma}_v\}$ and $\bb\,{=}\,(b_1,b_2)$ so that~\footnote{As in the dual-cycle case, $\bm{L}_\bb$ is defined by $\bb$ such that $\bm{L}_\bb$ has a nontrivial loop along the $i$-th direction if $b_i\,{\neq}\,0$.}
\begin{align} \label{eq:lx_param}
    e^{i \pi l'_e}\,{=}\,s^{\tbm,\bb} \tilde{\sigma}_v \tilde{\sigma}_{v'} \quad \textrm{with}\quad  \bs^{\tbm,\bb}\,{:=}\,e^{i \pi (\bl^\tbm\,{+}\,\bm{L}_\bb) },
\end{align}
where $s^{\tbm,\bb}_e$ is defined on the edge of the dual lattice.
Accordingly, the second factor is expressed as:
\begin{align}
     \hspace{-5pt} \sum_{ \substack{ \rd(\bl'-\bl^\tbm)=0} } \hspace{-10pt}  (1-p_x)^{2N - |\bl'|} p_x^{|\bl'|} &=\sum_{\bb} \frac{ Z[\bs^{\tilde{\bmm},\bb}, \beta_x] }{2(2\cosh \beta_x)^{2N}}.
\end{align}
Therefore, we diagonalized $\cE[\rho_0]$ with eigenvalue given in terms of the RBIM partition functions.

Since $\rho_Q = \frac{1}{4} \sum_{\ba}  \overline{\bm{Z}}_{\ba}^\vdagger \rho_0 \overline{\bm{Z}}_{\ba}^\dagger$, 
$\cE[\rho_Q]$ is diagonalized as 
\begin{align}
    \cE[\rho_Q] &=4 \sum_{\bmm, \tilde{\bmm},\ba} \rho_{\bmm \tilde{\bmm}}^{\ba, \ba} \cdot \opx \cdot \opz
    \nonumber \\
    p^i_{\bmm,\ba} &:= \frac{  {Z[\bs^{\bmm,\ba},\beta_i] }}{2 (2 \cosh \beta_i)^{{2N}}}, \quad \,\, \opi := \sum_\ba \frac{1}{4} p^i_{\bmm,\ba}.
\end{align}
It is straightforward to check that $\sum_{\bmm, \ba} p_{\bmm,\ba}^i\,{=}\,1$, thus 
$\cE[\rho_Q]$ is properly normalized.
Therefore, the entanglement entropy is given as
\begin{align} \label{eq:SQ}
    &S(Q) = - 2 \log 2 -\sum_{\bmm} 4 \opz \log \opz  - \sum_{\tilde{\bmm}} 4 \opx \log \opx
\end{align}
In the limit $\beta_i\,{\rightarrow}\,\infty$, the partition function of an RBIM is dominated by the contribution with the trivial frustration pattern with $\{m_p\,{=}\,1\}$ and $\{a_i\,{=}\,1\}$. This implies that  $\overline{p^i_\bmm}\,{\rightarrow}\, \delta_{ {\bmm}, {\bm{1}}}/4$. Therefore, we correctly recover $S(Q)\,{=}\,2 \log 2$ under the absence of errors.

\emph{Diagonalization II}.--- To get coherent information, we also need 
$S(\cE[\rho_{RQ}])$. 
In the combined system, we label the reference’s two qubits by $\bm{\alpha}\,{=}\,(\alpha_1,\alpha_2)$, the eigenvalues of $(X_1,X_2)$.
\begin{align}
    \rho_{RQ} &= \sum_{\bm{\alpha}, \bm{\alpha'}} \frac{1}{4} |\bm{\alpha} \rangle \langle \bm{\alpha'} | \otimes (\overline{\bm{Z}}_\balp^\vdagger \rho_0 \overline{\bm{Z}}_\balpp^\dagger)
\end{align}
Define $M_{\bm{\alpha}, \bm{\alpha'}}\,{:=}\,\overline{\bm{Z}}_\balp^\vdagger \rho_0 \overline{\bm{Z}}_\balpp^\dagger$.
In the basis $ \rho_{\bmm \tilde{\bmm}}^{\ba, \bap}$, each component of $\cE[M_{\bm{\alpha}, \bm{\alpha'}}]$ is given as
\begin{align}
    &\tr\big(\cE[M_{\bm{\alpha}, \bm{\alpha'}}] (\rho_{\bmm \tilde{\bmm}}^{\ba, \bap})^\dagger \big) = \sum_{\bl_x, \bl_z} \Big(\prod_{i\in x,z}   (1-p_i)_{\vphantom{i}}^{2N - |\bl_i|}  p_i^{|\bl_i|} \Big) \times \nonumber \\
    & \,\,\,\,\,\,\,\,\,  \tr( X_{\bl_x}^\vdagger Z_{\bl_z}^\vdagger M_{\bm{\alpha}, \bm{\alpha'}} Z_{\bl_z}^\dagger X_{\bl_x}^\dagger   X_{\bl^\tbm}^\vdagger  Z_{\bl^{\bmm}}^\vdagger M_{\bap,\ba}  Z_{\bl^{\bmm}}^\dagger  X_{\bl^\tbm}^\dagger   ) 
\end{align} 
For the trace not to vanish, two conditions are required:
\begin{align}
    (i) & \quad \rd_{\scriptscriptstyle{\perp}}(\bl_z-\bl^\bmm) \equiv \bm{0}, \quad  \rd(\bl_x-\bl^\tbm) \equiv \bm{0} \nonumber \\
    (ii) & \quad [\bl_z - \bl^\bmm] 
    \equiv \bap - \bm{\alpha'} \equiv \ba - \bm{\alpha}.
\end{align}
Unlike before, here the homotopy class of $\bm{l}_x\,{-}\,\bm{l}^\tbm$ influences a possible sign factor when commuting Pauli-$X$ loops.
Moving the Pauli-$X$ loops past $M_{\bm{\alpha},\bm{\alpha'}}$ introduces
\begin{align}
    &X_{\bl_x}^\dagger   X_{\bl^\tbm}^\vdagger  M_{\bm{\alpha},\bm{\alpha'}} X_{\bl^\tbm}^\dagger X_{\bl_x}^\vdagger  =   (\xi_\balp^{[\bl_x{-}\bl^\tbm]} )M_{\bm{\alpha}, \bm{\alpha'}} (\xi_\balpp^{[\bl_x{-}\bl^\tbm]} )^{-1} \nonumber \\
    &\qquad \qquad \xi_{\ba}^{[\bm{L}_\bb]} := e^{\mathrm{i} \pi (\ba \cdot \bb)}, \quad \ba \cdot \bb = \sum_i a_i b_i.
\end{align}
due to the non-trivial commutation relationship between logical operators $\overline{\bm{Z}}_\balp$ and $\overline{\bm{X}}_\bb$.
Accordingly, if we parametrize $\bl_x$ as in \eqnref{eq:lx_param}, we get
\begin{align} \label{eq:dft}
    &\tr\big(\cE[M_{\bm{\alpha}, \bm{\alpha'}}] (\rho_{\bmm \tilde{\bmm}}^{\ba, \bap})^\dagger \big) = p_{\bmm, \ba-\balp \vphantom{\tilde{\bmm}|\bm{\alpha}}}^z  \, p_{\tilde{\bmm}|\bm{\alpha}-\bm{\alpha'}}^x \cdot \delta_{\ba-\balp,\bap-\balpp}^{\vphantom{x}} \nonumber \\
    &\quad \textrm{where}\quad p_{\tilde{\bmm}|\bm{\alpha}}^x := \sum_\bb  \xi_\balp^{[\bm{L}_\bb]} p_{\tilde{\bmm},\bb}^x.
\end{align}
Therefore, the density matrix is
\begin{align}
    & \cE[\rho_{RQ}] = \sum_{\bm{\alpha}, \bm{\alpha'} } \sum_{\ba, \bmm, \tbm}  \frac{p_{\bmm, \ba \vphantom{\balpp}}^z   \,  p_{\tbm|\balp-\balpp}^x}{4}   | \bm{\alpha} \rangle \langle \bm{\alpha'} | \otimes \rho^{\ba + \balp, \ba+\balpp}_{\bmm, \tilde{\bmm}}.
\end{align}
It is block-diagonal in \((\bmm,\tilde{\bmm})\), and each \(16\times16\) block further breaks into four \(4\times4\) blocks because \(\rho_{\bmm,\tbm}^{\ba + \balp,\ba + \balpp}\) and \(\rho_{\bmm,\tbm}^{\bap + \balp,\bap + \balpp}\) are orthogonal whenever \(\ba\neq\bap\). Our task is thus to diagonalize the \(4\times4\) block

The density matrix is block-diagonal in $(\bmm, \tilde{\bmm})$, and each $16\,{\times}\,16$ block further breaks into four $4\,{\times}\,4$ blocks because $\rho^{\ba + \balp, \ba+\balpp}_{\bmm, \tilde{\bmm}}$ and $\rho^{\bap + \balp, \bap+\balpp}_{\bmm, \tilde{\bmm}}$ are orthogonal if $\ba\,{\neq}\,\bap$. Our task is thus to diagonalize the $4 \times 4$ block defined as 
\begin{align}
    B^\ba_{\bmm,\tbm} &:= \frac{p_{\bmm, {\ba}}^z}{4}  \sum_{\bm{\alpha}, \bm{\alpha'} }  p_{\tilde{\bmm}|\bm{\alpha}-\bm{\alpha'}}^x | \bm{\alpha} \rangle \langle \bm{\alpha'} | \otimes \rho^{\ba+\balp, \ba + \balpp}_{\bmm, \tbm}.
\end{align}
Eigenvectors of $B^\ba_{\bmm,\tbm}$ can be directly constructed as follows. Let $\bm{\beta}\,{=}\,(\beta_1,\beta_2)$ with $\beta_i\,{\in}\,\{0,1\}$. Define  
$v^{\bm{\beta}}\,{:=}\,\sum_{\balpp} \xi^{\bm{\beta}}_\balpp (| \balpp \rangle \otimes \overline{\bm{Z}}_{\ba+\balpp}^{\vphantom{\beta}} |\psi^\tc_0 \rangle)$. 
These are eigenvector with eigenvalue $p_{\bmm, \ba \vphantom{\beta}}^z p_{\tbm, \bm{\beta}}^x$:
\begin{align} \label{eq:basis}
    (B^\ba_{\bmm,\tbm}
    v^{\bm{\beta}})_\balp 
    &= p_{\bmm, {\ba}}^z \cdot \frac{1}{4}  \sum_\balpp \xi^{\bm{\beta}}_\balpp p_{\tilde{\bmm}|\balp - \balpp}^x \nonumber \\
    &= p_{\bmm \vphantom{\tilde{\bmm}}, {\ba}}^z \cdot  \xi_\balp^{\bm{\beta}} \cdot p_{\tilde{\bmm},{\bm{\beta}}}^x  = p_{\bmm \vphantom{\tilde{\bmm}}, {\ba}}^z p_{\tilde{\bmm},{\bm{\beta}}}^x  (v^{\bm{\beta}})_\balp,
\end{align}
since the sum is a two-dimensional inverse Fourier transform with period 2. (This idea can be generalized for $\mathbb{Z}_n$ case) The four eigenvalues in each $(\bmm,\tbm,\ba)$ block are $\{ p_{\bmm \vphantom{\tilde{\bmm}}, {\ba}}^z p_{\tilde{\bmm},{\bm{\beta}}}^x \}_{\beta_i\,{\in}\,\{0,1\}}$, corresponding to products of two RBIM partition functions at $\beta_x$ and $\beta_z$ with domain wall configurations $\ba$ and $\bb$. 
Finally, because our analysis allows for overlapping errors~\footnote{Even if we applied Pauli-$X$ and $Z$ error channels sequentially, both errors can be applied on some sites.} the basis we found  remains diagonal even if $X$ and $Z$ errors are correlated—only the probability distribution changes. For further details, see~\cite{SM}.

Thus, the entropy of $\cE[\rho_{RQ}]$ is evaluated as
\begin{align} \label{eq:SRQ}
    S( \cE[\rho_{RQ}] ) &= - \sum_{\bm{m},\ba \vphantom{\tbm,\bb}} \sum_{\tbm, \bb}
        \, p_{\bmm \vphantom{\tilde{\bmm}}, {\ba}}^z \,p_{\tilde{\bmm},{\bb}}^x \log \, p_{\bmm \vphantom{\tilde{\bmm}}, {\ba}}^z \, p_{\tilde{\bmm},{\bb}}^x
\end{align}
In the limit $\beta_{z,x}\,{\rightarrow}\,\infty$, both $p_{\bmm \vphantom{\tilde{\bmm}}, {\ba}}^z$ and $p_{\tilde{\bmm},{\bb}}^x$ are nonzero only if $(\bmm,\ba)$ and $(\tilde{\bmm},\bb)$ are trivial  configurations; in such a case, we obtain $S(R'Q')\,{=}\,0$ as expected.

\emph{Coherent Information}.--- 
Substituting \eqnref{eq:SQ} and \eqnref{eq:SRQ} into \eqnref{eq:coherentInfo} yields the coherent information of the decohered toric code:
\begin{align}  \label{eq:coh1}
    I_c^{\tc} 
    &= -2 \log 2 + \sum_{{\bmm},\ba,i } \qty( { p^i_{\bm{m},\ba\vphantom{\tilde{\bmm}}} } \Big[ \log  { p^i_{\bm{m},\ba\vphantom{\tilde{\bmm}}} } - \log  \opi \Big]  ) \nonumber \\
    &= 2 \log 2 + \sum_{\bmm,\ba,i}    p_{\bmm,\ba}^i \log( \frac{ {Z[\bs^{\bmm,\ba},\beta_i]}}{ \sum_{\bap} Z[\bs^{\bmm,\bap},\beta_i] } )
\end{align}
where $i\in\{x,z\}$ and $(\bmm,\ba)$ labels the RBIM bond-frustration pattern.
With this analytic expression, we can rigorously determine the toric code’s information capacity under Pauli errors in the thermodynamic limit.

Let $\beta_c$ be the RBIM’s Nishimori-line critical point~\cite{Nishimori, Nishimori2}, with $p_c=0.1094$~\cite{NishimoriPoint}. 
To facilitate the analysis, define $F_{\bmm,\ba}^i\,{:=}\,{-}\log p^i_{\bmm,\ba}$, which is the free energy of a RBIM. The difference $\Delta^i_{\bmm,\ba}\,{:=}\,F^i_{\bmm,\ba}\,{-}\,F^i_{\bmm,\bm{0}}$ corresponds to the free energy cost of inserting a domain wall $\ba$ from the configuration $(\bmm,\bm{0})$. 
Then the second term in \eqnref{eq:coh1} separates as $I_c^\tc=2\log 2 - A_x - A_z$, where 
\begin{align} \label{eq:CI_final}
    A_i = \sum_{\bmm}  p_{ \bmm,\bm{0}}^i \sum_\ba \qty[ e^{-\Delta^i_{\bmm,\ba}  } \log  \frac{ \,\sum_{\bap} e^{-\Delta^i_{\bmm,\bap}  } }{  e^{-\Delta^i_{\bmm,\ba}  }  }  ].
\end{align} 
We examine $A_i$ in the thermodynamic limit.

\vspace{2pt} \noindent {\bf (1)} $\beta_i\,{>}\,\beta_c$: 
Most bond configurations (in the disorder ensemble of the RBIM) are long-range ordered. Thus, the denominator inside $\log(\cdots)$ in \eqnref{eq:CI_final} is dominated by $\ba\,{=}\,\bm{0}$ (no domain wall). 
When a given configuration is long-range ordered, the cost of domain wall insertion scales with system size, $|\Delta^x_{\bmm,\ba}|\,{\geq}\,cL \delta_{\ba,\bm{0}}$ for some non-zero constant $c$ and $L\,{=}\,\sqrt{N}$. 
Within the disorder ensemble, the fraction of paramagnetic bond configurations vanishes as the system size increases, as detailed in \cite{SM}. 
With this premise, one can establish that $\lim_{N \rightarrow \infty} A_i\,{=}\,0$.
Therefore, if $\beta_{x,z} \,{>}\,\beta_c$ ($p_{x,z}\,{<}\,p_c$), $I_c^\tc \rightarrow 2 \log 2$ and two qubits of decodable quantum information persist.

\vspace{2pt} \noindent  {\bf (2)} $\beta_i\,{<}\,\beta_c$: 
Most bond configurations are paramagnetic and $\Delta^z_{\bmm,\ba}\,{\rightarrow}\,0$ in the thermodynamic limit since domain walls are freely fluctuating in the paramagnetic phase~\footnote{This is expected since disorder operators' correlation functions decay exponentially with their perimeter size in the disordered phase.}. Accordingly, $A_i\,{\rightarrow}\,2 \log 2$, see \cite{SM}. 
Without loss of generality, if $\beta_x\,{<}\,\beta_c$ and $\beta_z\,{>}\,\beta_c$, $A_x\,{=}\,2 \log 2$ and $A_z\,{=}\,0$ and we get $I_c^\tc = 0$, implying that there remain two bits of classical information that can be restored, which corresponds to the eigenvalues of $(\overline{\bm{X}}_1,\overline{\bm{X}}_2)$. In the doubled Hilbert space formalism~\cite{LeeSPT2022, bao2023mixedstate}, this corresponds to the phase where two copies of $\mathbb{Z}_2$ topological order condense into a single $\mathbb{Z}_2$ topological order~\cite{Lee2023PRXQ, Fan2023}.
On the other hand, if both $\beta_{x,z}\,{<}\,\beta_c$, $I_c \approx - 2 \log 2$, there remains neither quantum nor classical information that can be decoded. This corresponds to the trivial topological order in the doubled Hilbert space.

It is instructive to compare with the Renyi-2 coherent information~\cite{Fan2023}, tied to the free energy cost $\Delta_{\bm{0},\ba}$ of a ferromagnetic Ising model at $\tilde{\beta}_i=- \tfrac12 \log(1-2p_i)$~\footnote{At $p=0$, $\tilde{\beta}=0$ while $\beta=\infty$.}:
\begin{align}
    {I^{\tc,(2)}_c} = \sum_{i={x,y}} \log\bigg( \sum_{\ba} e^{-\Delta^{i}_{\bm{0},\ba}} \bigg) - 2 \log 2,
\end{align}
which transitions at $p_c'=0.178$~\cite{Lee2023PRXQ, Fan2023}.  

To illustrate the advantage of storing information in the toric code, we calculate the coherent information for two raw qubits under the same channel:
\begin{align}
    I_c^\textrm{raw} &= 2 \Big(\log 2 - \hspace{-5pt} \sum_{i \in \{x,z\} } \hspace{-3pt}  H_2(p_i) \Big).
\end{align}
where $H_2(p)\,{:=}\,\,{-}\,p \log p\,{-}\,(1-p)\log(1-p)$. 
As shown in Fig.~\ref{fig:2}(a,b), below $p_c$, $I_c^\tc\,{=}\,2\log2$ but $I_c^\textrm{raw}$ decreases. If $p_i\,{>}\,p_c$, $I_c^\textrm{raw}$ can exceed $I_c^\tc$, removing any benefit from encoding. We remark that along $p_x=p_z$, $p_c=0.1094$ is close to $p=0.1100$ where $I_c^\textrm{raw}=0$.

\emph{Relative Entropy.}--- With a diagonalized decohered density matrix, it is straightforward to evaluate a quantum relative entropy~\cite{Relative} between two decohered toric code states, each of which is initialized at different logical states. Consider two initial states $\rho_0$ and $\rho_1\,{=}\,\overline{\bm{Z}}_\ba \rho_0 \overline{\bm{Z}}_\ba$. If $\rho_n'\,{=}\,\cE[\rho\vphantom{'}_n]$, the relative entropy between two decohered states is given as
\begin{align}
    D( \rho_0' \Vert \rho_1' )
    &:= \tr\big( \rho_0' (\log \rho_0' - \log \rho_1') \big) \nonumber \\
    &=  \sum_{\bmm,\bap} p^x_{\bmm,\bap} (F^x_{\bmm,\ba\bap} - F^x_{\bmm,\bap})
    = \langle \Delta F_{\bm{\ba}}  \rangle
\end{align}
which is the disorder-averaged free energy of a domain wall configuration $\ba$ in the RBIM along the Nishimori line at $\beta_x$ (see \cite{SM}). 
If $\beta_x\,{>}\,\beta_c$ ($p_x\,{<}\,p_c$), the system is long-range ordered on average, and $\langle \Delta F_\ba \rangle\,{\sim}\,{\cal O}(L)$.  
As the relative quantum entropy signifies the distinguishability of two states, the linear scaling of $\langle \Delta F_\ba \rangle$ implies that two different logical states are well-distinguishable even after local decoherence in the thermodynamic limit.  If $\beta_x\,{<}\,\beta_c$, $\langle \Delta F_\ba \rangle\,{\sim}\,0$ in the thermodynamic limit and two different logical states would be indistinguishable. 

We note a critical distinction from coherent information: relative entropy (disorder averaged free energy) would exhibit $\cO(L)$
scaling even if a constant fraction of the disorder ensemble were in the paramagnetic phase (see \cite{SM}), which would lead to coherent information being strictly smaller than $2 \log 2$ in the thermodynamic limit. Consequently, relative entropy or ``free energy'' of a domain wall emerges as a less refined measure for quantifying decodable quantum information.

\emph{Failure of free energy criteria}.--- Given syndrome observations $(\bmm,\tbm)$,  we can obtain probabilities for different values of $(\ba, \bb)$ by calculating the partition function for the associated random bond model. In the maximum entropy decoder, we correct errors based on these probabilities. When we have an error $(\bmm,\tbm,\ba,\bb)$, the decoding is successful only if we infer the same error class, whose probability is given as
\begin{align} \label{eq:condition1}
   P^{\textrm{suc.}}_{\bmm,\tbm,\ba,\bb} = \frac{ P_{\bmm,\tbm,\ba,\bb} }{ \sum_{\ba',\bb'}P_{\bmm,\tbm,\ba',\bb'} }.
\end{align}
where $P_{\bmm,\tbm,\ba,\bb} = p^z_{\bmm,\bm{a}} p^x_{\tbm,\bm{b}}$ is the probability of getting this error under independent $Z$ and $X$ errors.
The average success probability is then given as
\begin{align}
\overline{ P^{\textrm{suc.}} } = \sum_{\bmm,\tbm, \ba, \bb } P_{\bmm,\tbm,\ba,\bb}  \cdot P^{\textrm{suc.}}_{\bmm,\tbm,\ba,\bb}.
\end{align}
The lower bound of this success probability is determined by the coherent information; by Jensen's inequality, 
\begin{align}
     \log \overline{ P^\textrm{suc.} }  \geq \overline{ \log P^\textrm{suc.}  } = I_c - k \log 2.
\end{align}
Therefore, $e^{I_c - k \log 2}\,{\leq}\,\overline{ P^{\textrm{suc.}} }$ and  $I_c = k \log 2$ is a sufficient condition for the maximum entropy decoder to success always. The divergence of domain wall free energy provides only a lower bound on the failure rate~(See \cite{SM}), implying that the actual failure rate can be larger. On the other hand, coherent information provides a lower bound on the success rate and can be used as a useful metric within the correctable regime.

\emph{Conclusion}.--- 
By exactly diagonalizing the toric code plus its reference under decoherence, we derived a closed-form expression for the toric code’s coherent information.
Our formula links directly to the random-bond Ising model’s free energy, revealing the coherent-information transition and fundamental error threshold under Pauli errors.
We also showed that free energy-based thresholds can miss crucial details, whereas coherent information captures the true limit.

The observation that exact diagonalization yields the same statistical model as maximum-entropy decoding suggests that the fundamental thresholds of other models may likewise be determined by the coherent information form in \eqnref{eq:coh1}.
In addition, the formalism developed in this work directly extends to a decoherence model with correlated $X$ and $Z$ errors~\cite{SM}, as well as for the family of Calderbank-Shor-Steane codes~\cite{PhysRevA.111.032402} and their $\mathbb{Z}_n$ generalizations.

\acknowledgments
We thank Matthew F.A. Fisher, Ehud Altman, Ruihua Fan, Sajant Anand, Soonwon Choi, and Jeongwan Haah, for fruitful discussions and comments. 
The work is supported by the Simons Investigator Award and a faculty startup grant at the University of Illinois, Urbana-Champaign.

\bibliography{ref}

@article{Chen:2023tfg,
    author = "Chen, Edward H. and others",
    title = "{Nishimori transition across the error threshold for constant-depth quantum circuits}",
    eprint = "2309.02863",
    archivePrefix = "arXiv",
    primaryClass = "quant-ph",
    doi = "10.1038/s41567-024-02696-6",
    journal = "Nature Phys.",
    volume = "21",
    number = "1",
    pages = "161--167",
    year = "2025"
}

@article{PhysRevA.111.032402,
  title = {Coherent information for Calderbank-Shor-Steane codes under decoherence},
  author = {Niwa, Ryotaro and Lee, Jong Yeon},
  journal = {Phys. Rev. A},
  volume = {111},
  issue = {3},
  pages = {032402},
  numpages = {13},
  year = {2025},
  month = {Mar},
  publisher = {American Physical Society},
  doi = {10.1103/PhysRevA.111.032402},
  url = {https://link.aps.org/doi/10.1103/PhysRevA.111.032402}
}

@misc{SM,
title={See Supplemental Material, which includes Refs.[20, 40-42].}
}

@article{PhysRevLett.131.200201,
  title = {Nishimori's Cat: Stable Long-Range Entanglement from Finite-Depth Unitaries and Weak Measurements},
  author = {Zhu, Guo-Yi and Tantivasadakarn, Nathanan and Vishwanath, Ashvin and Trebst, Simon and Verresen, Ruben},
  journal = {Phys. Rev. Lett.},
  volume = {131},
  issue = {20},
  pages = {200201},
  numpages = {9},
  year = {2023},
  month = {Nov},
  publisher = {American Physical Society},
  doi = {10.1103/PhysRevLett.131.200201},
  url = {https://link.aps.org/doi/10.1103/PhysRevLett.131.200201}
}

@article{PhysRevLett.104.050504,
  title = {Fast Decoders for Topological Quantum Codes},
  author = {Duclos-Cianci, Guillaume and Poulin, David},
  journal = {Phys. Rev. Lett.},
  volume = {104},
  issue = {5},
  pages = {050504},
  numpages = {4},
  year = {2010},
  month = {Feb},
  publisher = {American Physical Society},
  doi = {10.1103/PhysRevLett.104.050504},
  url = {https://link.aps.org/doi/10.1103/PhysRevLett.104.050504}
}

@misc{kovalev2014spin,
      title={Spin glass reflection of the decoding transition for quantum error correcting codes}, 
      author={Alexey A. Kovalev and Leonid P. Pryadko},
      year={2014},
      eprint={1311.7688},
      archivePrefix={arXiv},
      primaryClass={quant-ph}
}

@misc{sang2023mixedstate,
      title={Mixed-state Quantum Phases: Renormalization and Quantum Error Correction}, 
      author={Shengqi Sang and Yijian Zou and Timothy H. Hsieh},
      year={2023},
      eprint={2310.08639},
      archivePrefix={arXiv},
      primaryClass={quant-ph}
}

@misc{colmenarez2024accurate,
      title={Accurate optimal quantum error correction thresholds from coherent information}, 
      author={Luis Colmenarez and Ze-Min Huang and Sebastian Diehl and Markus Müller},
      year={2024},
      eprint={2312.06664},
      archivePrefix={arXiv},
      primaryClass={quant-ph}
}

@article{HastingsSeparability2011,
  title = {Topological Order at Nonzero Temperature},
  author = {Hastings, Matthew B.},
  journal = {Phys. Rev. Lett.},
  volume = {107},
  issue = {21},
  pages = {210501},
  numpages = {5},
  year = {2011},
  month = {Nov},
  publisher = {American Physical Society},
  doi = {10.1103/PhysRevLett.107.210501},
  url = {https://link.aps.org/doi/10.1103/PhysRevLett.107.210501}
}

@misc{chen2023separability,
      title={Separability transitions in topological states induced by local decoherence}, 
      author={Yu-Hsueh Chen and Tarun Grover},
      year={2023},
      eprint={2309.11879},
      archivePrefix={arXiv},
      primaryClass={quant-ph}
}

@misc{SurfaceCode1998,
      title={Quantum codes on a lattice with boundary}, 
      author={S. B. Bravyi and A. Yu. Kitaev},
      year={1998},
      eprint={quant-ph/9811052},
      archivePrefix={arXiv},
      primaryClass={quant-ph}
}

@article{Knill,
author = {Emanuel Knill  and Raymond Laflamme  and Wojciech H. Zurek },
title = {Resilient Quantum Computation},
journal = {Science},
volume = {279},
number = {5349},
pages = {342-345},
year = {1998},
doi = {10.1126/science.279.5349.342},
URL = {https://www.science.org/doi/abs/10.1126/science.279.5349.342}}

@article{Kitaev_1997,
doi = {10.1070/RM1997v052n06ABEH002155},
url = {https://dx.doi.org/10.1070/RM1997v052n06ABEH002155},
year = {1997},
month = {dec},
publisher = {},
volume = {52},
number = {6},
pages = {1191},
author = {A Yu Kitaev},
title = {Quantum computations: algorithms and error correction},
journal = {Russian Mathematical Surveys}
}

@inproceedings{Aharonov1997,
author = {Aharonov, D. and Ben-Or, M.},
title = {Fault-Tolerant Quantum Computation with Constant Error},
year = {1997},
isbn = {0897918886},
publisher = {Association for Computing Machinery},
address = {New York, NY, USA},
url = {https://doi.org/10.1145/258533.258579},
doi = {10.1145/258533.258579},
booktitle = {Proceedings of the Twenty-Ninth Annual ACM Symposium on Theory of Computing},
pages = {176–188},
numpages = {13},
location = {El Paso, Texas, USA},
series = {STOC '97}
}

@article{Gottesman1998,
  title = {Theory of fault-tolerant quantum computation},
  author = {Gottesman, Daniel},
  journal = {Phys. Rev. A},
  volume = {57},
  issue = {1},
  pages = {127--137},
  numpages = {0},
  year = {1998},
  month = {Jan},
  publisher = {American Physical Society},
  doi = {10.1103/PhysRevA.57.127},
  url = {https://link.aps.org/doi/10.1103/PhysRevA.57.127}
}

@article{Relative,
    author = {Lieb, Elliott H. and Ruskai, Mary Beth},
    title = "{Proof of the strong subadditivity of quantum‐mechanical entropy}",
    journal = {Journal of Mathematical Physics},
    volume = {14},
    number = {12},
    pages = {1938-1941},
    year = {2003},
    month = {11},
    issn = {0022-2488},
    doi = {10.1063/1.1666274},
    url = {https://doi.org/10.1063/1.1666274},
}

@article{Shor1995,
  title = {Scheme for reducing decoherence in quantum computer memory},
  author = {Shor, Peter W.},
  journal = {Phys. Rev. A},
  volume = {52},
  issue = {4},
  pages = {R2493--R2496},
  numpages = {0},
  year = {1995},
  month = {Oct},
  publisher = {American Physical Society},
  doi = {10.1103/PhysRevA.52.R2493},
  url = {https://link.aps.org/doi/10.1103/PhysRevA.52.R2493}
}

@article{Bombin2012,
  title = {Strong Resilience of Topological Codes to Depolarization},
  author = {Bombin, H. and Andrist, Ruben S. and Ohzeki, Masayuki and Katzgraber, Helmut G. and Martin-Delgado, M. A.},
  journal = {Phys. Rev. X},
  volume = {2},
  issue = {2},
  pages = {021004},
  numpages = {10},
  year = {2012},
  month = {Apr},
  publisher = {American Physical Society},
  doi = {10.1103/PhysRevX.2.021004},
  url = {https://link.aps.org/doi/10.1103/PhysRevX.2.021004}
}

@article{Katzgraber2009,
  title = {Error Threshold for Color Codes and Random Three-Body Ising Models},
  author = {Katzgraber, Helmut G. and Bombin, H. and Martin-Delgado, M. A.},
  journal = {Phys. Rev. Lett.},
  volume = {103},
  issue = {9},
  pages = {090501},
  numpages = {4},
  year = {2009},
  month = {Aug},
  publisher = {American Physical Society},
  doi = {10.1103/PhysRevLett.103.090501},
  url = {https://link.aps.org/doi/10.1103/PhysRevLett.103.090501}
}

@article{noclone,
	author = {Wootters, W. K. and Zurek, W. H.},
	date = {1982/10/01},
	doi = {10.1038/299802a0},
	id = {Wootters1982},
	isbn = {1476-4687},
	journal = {Nature},
	number = {5886},
	pages = {802--803},
	title = {A single quantum cannot be cloned},
	url = {https://doi.org/10.1038/299802a0},
	volume = {299},
	year = {1982}}

@article{shannon,
  author = {Shannon, Claude Elwood},
  journal = {The Bell System Technical Journal},
  pages = {379--423},
  title = {A Mathematical Theory of Communication},
  url = {http://plan9.bell-labs.com/cm/ms/what/shannonday/shannon1948.pdf},
  urldate = {2003-04-22},
  volume = 27,
  year = 1948
}

@article{Kubica2018,
  title = {Three-Dimensional Color Code Thresholds via Statistical-Mechanical Mapping},
  author = {Kubica, Aleksander and Beverland, Michael E. and Brand\~ao, Fernando and Preskill, John and Svore, Krysta M.},
  journal = {Phys. Rev. Lett.},
  volume = {120},
  issue = {18},
  pages = {180501},
  numpages = {6},
  year = {2018},
  month = {May},
  publisher = {American Physical Society},
  doi = {10.1103/PhysRevLett.120.180501},
  url = {https://link.aps.org/doi/10.1103/PhysRevLett.120.180501}
}

@ARTICLE{Chubb2021,
       author = {{Chubb}, Christopher T. and {Flammia}, Steven T.},
        title = "{Statistical mechanical models for quantum codes with correlated noise}",
      journal = {Annales de l'Institut Henri Poincare D},
         year = 2021,
        month = jan,
       volume = {8},
       number = {2},
        pages = {269-321},
          doi = {10.4171/AIHPD/105},
archivePrefix = {arXiv},
       eprint = {1809.10704},
 primaryClass = {quant-ph},
       adsurl = {https://ui.adsabs.harvard.edu/abs/2021AIHPD...8..269C}
}

@article{Lee2023PRXQ,
  title = {Quantum Criticality Under Decoherence or Weak Measurement},
  author = {Lee, Jong Yeon and Jian, Chao-Ming and Xu, Cenke},
  journal = {PRX Quantum},
  volume = {4},
  issue = {3},
  pages = {030317},
  numpages = {20},
  year = {2023},
  month = {Aug},
  publisher = {American Physical Society},
  doi = {10.1103/PRXQuantum.4.030317},
  url = {https://link.aps.org/doi/10.1103/PRXQuantum.4.030317}
}

@ARTICLE{LeeNishimori2022,
       author = {{Lee}, Jong Yeon and {Ji}, Wenjie and {Bi}, Zhen and {Fisher}, Matthew P.~A.},
        title = "{Decoding Measurement-Prepared Quantum Phases and Transitions: from Ising model to gauge theory, and beyond}",
      journal = {arXiv e-prints},
         year = 2022,
        month = aug,
          eid = {arXiv:2208.11699},
        pages = {arXiv:2208.11699},
          doi = {10.48550/arXiv.2208.11699},
archivePrefix = {arXiv},
       eprint = {2208.11699},
 primaryClass = {cond-mat.str-el},
       adsurl = {https://ui.adsabs.harvard.edu/abs/2022arXiv220811699L}
}

@ARTICLE{Fan2023,
       author = {{Fan}, Ruihua and {Bao}, Yimu and {Altman}, Ehud and {Vishwanath}, Ashvin},
        title = "{Diagnostics of mixed-state topological order and breakdown of quantum memory}",
      journal = {arXiv e-prints},
         year = 2023,
        month = jan,
          eid = {arXiv:2301.05689},
        pages = {arXiv:2301.05689},
          doi = {10.48550/arXiv.2301.05689},
archivePrefix = {arXiv},
       eprint = {2301.05689},
 primaryClass = {quant-ph},
       adsurl = {https://ui.adsabs.harvard.edu/abs/2023arXiv230105689F}
}

@misc{bao2023mixedstate,
      title={Mixed-state topological order and the errorfield double formulation of decoherence-induced transitions}, 
      author={Yimu Bao and Ruihua Fan and Ashvin Vishwanath and Ehud Altman},
      year={2023},
      eprint={2301.05687},
      archivePrefix={arXiv},
      primaryClass={quant-ph}
}

@ARTICLE{LeeSPT2022,
       author = {{Lee}, Jong Yeon and {You}, Yi-Zhuang and {Xu}, Cenke},
        title = "{Symmetry protected topological phases under decoherence}",
      journal = {arXiv e-prints},
         year = 2022,
        month = oct,
          eid = {arXiv:2210.16323},
        pages = {arXiv:2210.16323},
          doi = {10.48550/arXiv.2210.16323},
archivePrefix = {arXiv},
       eprint = {2210.16323},
 primaryClass = {cond-mat.str-el},
       adsurl = {https://ui.adsabs.harvard.edu/abs/2022arXiv221016323L}
}

@article{coherentInfo2,
	author = {Horodecki, Micha{\l} and Oppenheim, Jonathan and Winter, Andreas},
	date = {2007/01/01},
	doi = {10.1007/s00220-006-0118-x},
	id = {Horodecki2007},
	isbn = {1432-0916},
	journal = {Communications in Mathematical Physics},
	number = {1},
	pages = {107--136},
	title = {Quantum State Merging and Negative Information},
	url = {https://doi.org/10.1007/s00220-006-0118-x},
	volume = {269},
	year = {2007}}

@article{coherentInfo1,
  title = {Quantum data processing and error correction},
  author = {Schumacher, Benjamin and Nielsen, M. A.},
  journal = {Phys. Rev. A},
  volume = {54},
  issue = {4},
  pages = {2629--2635},
  numpages = {0},
  year = {1996},
  month = {Oct},
  publisher = {American Physical Society},
  doi = {10.1103/PhysRevA.54.2629},
  url = {https://link.aps.org/doi/10.1103/PhysRevA.54.2629}
}

@article{NishimoriPoint,
  title = {Universality Class of the Nishimori Point in the 2D $\ifmmode\pm\else\textpm\fi{}\mathit{J}$ Random-Bond Ising Model},
  author = {Honecker, A. and Picco, M. and Pujol, P.},
  journal = {Phys. Rev. Lett.},
  volume = {87},
  issue = {4},
  pages = {047201},
  numpages = {4},
  year = {2001},
  month = {Jul},
  publisher = {American Physical Society},
  doi = {10.1103/PhysRevLett.87.047201},
  url = {https://link.aps.org/doi/10.1103/PhysRevLett.87.047201}
}

@article{Nishimori2,
	author = {Nishimori ,Hidetoshi},
	doi = {10.1143/JPSJ.55.3305},
	journal = {Journal of the Physical Society of Japan},
	number = {10},
	pages = {3305-3307},
	title = {Geometry-Induced Phase Transition in the $\pm$J Ising Model},
	url = {https://doi.org/10.1143/JPSJ.55.3305},
	volume = {55},
	year = {1986}}

@article{Nishimori,
    author = {Nishimori, Hidetoshi},
    title = "{Internal Energy, Specific Heat and Correlation Function of the Bond-Random Ising Model}",
    journal = {Progress of Theoretical Physics},
    volume = {66},
    number = {4},
    pages = {1169-1181},
    year = {1981},
    month = {10},
    issn = {0033-068X},
    doi = {10.1143/PTP.66.1169},
    url = {https://doi.org/10.1143/PTP.66.1169},
}

@article{Dennis2002TQM,
	author = {Dennis,Eric and Kitaev,Alexei and Landahl,Andrew and Preskill,John},
	doi = {10.1063/1.1499754},
	journal = {Journal of Mathematical Physics},
	number = {9},
	pages = {4452-4505},
	title = {Topological quantum memory},
	url = {https://doi.org/10.1063/1.1499754},
	volume = {43},
	year = {2002},
	}

@article{MerzChalker2002,
  title = {Two-dimensional random-bond Ising model, free fermions, and the network model},
  author = {Merz, F. and Chalker, J. T.},
  journal = {Phys. Rev. B},
  volume = {65},
  issue = {5},
  pages = {054425},
  numpages = {18},
  year = {2002},
  month = {Jan},
  publisher = {American Physical Society},
  doi = {10.1103/PhysRevB.65.054425},
  url = {https://link.aps.org/doi/10.1103/PhysRevB.65.054425}
}

@article{WangPreskill2003,
	author = {Chenyang Wang and Jim Harrington and John Preskill},
	doi = {https://doi.org/10.1016/S0003-4916(02)00019-2},
	issn = {0003-4916},
	journal = {Annals of Physics},
	number = {1},
	pages = {31-58},
	title = {Confinement-Higgs transition in a disordered gauge theory and the accuracy threshold for quantum memory},
	url = {https://www.sciencedirect.com/science/article/pii/S0003491602000192},
	volume = {303},
	year = {2003},
	}

\newpage

\onecolumngrid


\makeatletter
\def\l@subsection#1#2{}
\def\l@subsubsection#1#2{}
\makeatother

\setcounter{equation}{0}
\setcounter{figure}{0}
\setcounter{table}{0}

\makeatletter
\renewcommand{\theequation}{S\arabic{equation}}
\renewcommand{\thefigure}{S\arabic{figure}}
\setcounter{subsection}{0}

\begin{center}
    \textbf{\large Supplementary Material for \\ \vspace{7pt}
    ``Exact Calculations of Coherent Information for Toric Codes under Decoherence: Identifying the Fundamental Error Threshold''}

    \vskip6mm
    
    {\noindent \normalsize  Jong Yeon Lee}

    \vskip3mm

    \noindent \normalsize  \emph{Department of Physics and Institute of Condensed Matter Theory} 
    
    \emph{University of Illinois at Urbana-Champaign, Urbana, Illinois 61801, USA}

    \end{center}


\section{Review on Random Bond Isind Model} \label{app:RBIM}

Consider a lattice with edges $(e)$ and vertices $(v)$. For a given bond configuration $\bb\,{=}\,\{ b_e \}$ with $b_e\,{\in}\,\{1,-1\}$, a random bond Ising model partition function at inverse temperature $\beta$ is defined as
\begin{align}
    Z_\textrm{RBIM}[\bb,\beta] := \sum_{\{\sigma\}} e^{-\beta \sum_{\langle v,v'\rangle} b_{v,v'} \sigma_v \sigma_{v'} }.
\end{align}
One crucial property of the RBIM is that its partition function is invariant under the gauge transformation of the bond configuration $\bb$ defined as follows. Consider a set of values defined on vertices $\bm{t}\,{=}\,\{ t_v \}$ with $t_v\,{\in}\{1,-1\}$. Define the transformation $\bb'\,{=}\,\cG[\bb,\bm{t}]$ such that $b'_{e}\,{=}\,t_v b_e t_{v'}$ for $e\,{=}\,(v,v')$. Then for any $\bm{t}$ with $\bb'\,{=}\,\cG[\bb,\bm{t}]$,
\begin{align}
    Z_\textrm{RBIM}[\bb,\beta] = Z_\textrm{RBIM}[\bb',\beta].
\end{align}
This is because the partition function for $\bb'$ is related to $\bb$ via a change of variables for its spin degrees of freedom.
Therefore, the partition function only depends on the equivalence class of $\bb$, which can be labeled by the gauge-invariant quantities $\bmm\,{=}\,\{m_p\}$ 
and $\ba\,{=}\,(a_1,a_2)$ defined as
\begin{align}
    m_p := \prod_{e \in p} b_e, \quad e^{i\pi a_i} := \prod_{e \in C_i} b_e
\end{align}
where $C_1$ and $C_2$ are particular cycles along $x$ and $y$ directions, respectively, see Fig.~2(a) in the main text.

Let $P(\bb)$ be the probability for a bond configuration to be $\bb$. If each bond is independent and ferromagnetic with probability $1-p$, then the probability distribution $P(\bb)$ is expressed as
\begin{align} \label{eq:Pb}
    P(\bb) = \prod_{\be} \sqrt{(1-p)^{1+b_e} p^{1-b_e} }= \frac{e^{\beta_p \sum_e b_e} }{(2\cosh \beta_p)^{2N}},
\end{align}
where $\beta_p = \tanh^{-1}(1-2p)$.



\section{Nishimori Line, Free Energy, and Variance across Disorder Ensemble}\label{app:num_analysis}

Assuming $P(\bb)$ in \eqnref{eq:Pb},
the disorder averaged free energy is given as 
\begin{align}
    \beta \overline{F} &:= - \sum_\bb P(\bb) \ln Z_\textrm{RBIM}[\bb,\beta] \nonumber \\
     &= - \sum_\bb \frac{e^{ \beta_p \sum_e b_e}}{(2\cosh \beta_p )^{2N}} \ln Z_\textrm{RBIM}[\bb,\beta] \nonumber \\
    &= - \hspace{-10pt} \sum_{ \bb' = \cG[\bb,t]} \frac{e^{ \beta_p \sum_{\langle v,v' \rangle } b'_{v,v'} t_v t_{v'}}}{ (2\cosh \beta_p)^{2N}} \ln Z_\textrm{RBIM}[\bb',\beta] \nonumber \\
    &= - \hspace{-1pt} \sum_{\bm{t}, \bb' } \frac{e^{ \beta_p \sum_{\langle v,v' \rangle } b'_{v,v'} t_v t_{v'}}}{2^{N}(2\cosh \beta_p)^{2N}} \ln Z_\textrm{RBIM}[\bb',\beta] \nonumber \\
    &= -  \sum_{\bb' } \frac{ Z_\textrm{RBIM}[\bb',\beta_p ] }{2^{N}(2\cosh \beta_p)^{2N}} \ln Z_\textrm{RBIM}[\bb',\beta]
\end{align}
where the line above $F$ indicates it is disorder averaged. In the third line, we used the change of variable from $\bb$ to $\bb'$ by the gauge transformation $\cG[\cdot,\bm{t}]$. 
In the fourth line, we use the fact that the summation over an auxiliary variable $\bm{t}$ introduces a multiplicative factor $1/2^N$.

The Nishimori condition, as established in the literature \cite{Nishimori,Nishimori2}, identifies a unique line in the phase diagram characterized by the equality $\beta_p = \beta$, which gives $p = \frac{1 - \tanh(\beta)}{2}$, directly linking the disorder fraction in the system to the thermal energy scale governed by $\beta$.

Previous numerical simulation has elucidated a critical phase transition point at $\beta_c\,{=}\, 1.048$~\cite{NishimoriPoint, MerzChalker2002} at which the system transitions from a state exhibiting long-range ferromagnetic order to a paramagnetic phase. As expected, it corresponds to a phase transition threshold at a lower critical temperature (or higher $\beta$) compared to the conventional ferromagnetic Ising model, which possesses a critical inverse temperature of $\beta_c^\textrm{FIM}\,{=}\,0.371$.

In the context of a disorder-averaged system, the characterization of a long-range ordered phase requires careful consideration. Specifically, the system may present a scenario where a constant fraction (${<}\,1$) of the bond configurations manifests long-range order, while the remaining configurations are paramagnetic. 
While one could argue that $\lim_{N \rightarrow \infty} f = 0$ based on the self-averaging property observed in many disordered systems, such an assertion does not universally hold. In such a case, the disorder-averaged order parameter would suggest the presence of long-range order, even if the system is essentially bifurcated between ferromagnetic and paramagnetic states. This observation underscores the limitation of a naive averaged order parameter (or averaged free energy) in accurately reflecting the system's state under varying degrees of disorder.

To discern whether the majority of bond configurations truly exhibit long-range order, the order parameter fluctuation across different bond configurations should be examined. For a given bond configuration $\bb$, the order parameter is defined as
\begin{align}
    \langle O \rangle_\bb := \frac{1}{Z[\bb_i,\beta]} \sum_{\{ \sigma \} } O(\{ \sigma \} ) e^{- \beta \sum b_e \sigma_v \sigma_{v'}}.
\end{align}
Assume that $O$ is normalized in such a way that $O\,{\in}\,[0,1]$; it takes a finite value in the long-range ordered phase, and zero in the paramagnetic phase. The variance across bond configurations is defined as
\begin{align}
    \chi := \overline{ \langle O \rangle^2_\bb }^\bb - \Big( \overline{\langle O \rangle_\bb}^\bb \Big)^2,
\end{align}
where $\overline{(\,\cdot\,)}^\bb\,{:=}\,\sum_\bb P(\bb) (\cdot)$ is disorder average. If $O$ is magnetization, $\chi$ is related to the magnetic susceptibility. If $\chi\,{\sim}\,\cO(1)$, then it implies that the constant fraction of bond configurations is paramagnetic. On the other hand, if $\chi\,{\sim}\,\cO(1/N)$, then the paramagnetic fraction is $1-\cO(1/N)$, vanishing in the thermodynamic limit.

\begin{figure}[!t]
    \includegraphics[width=0.7\columnwidth]{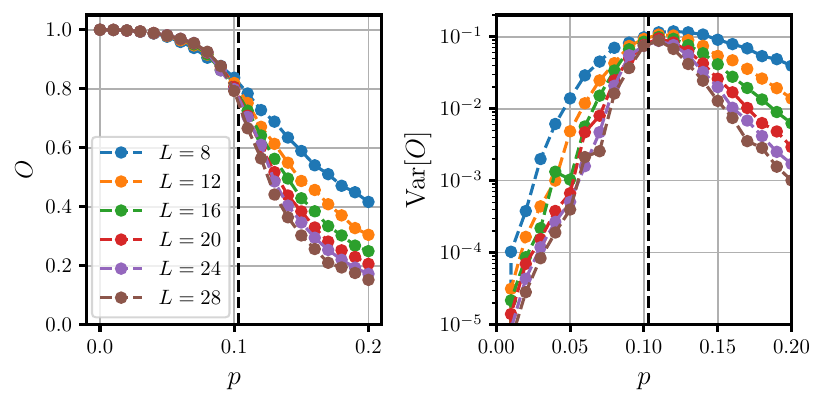}
    \caption{{\bf Random bond Ising model at $T=0$.} The plot shows {\bf (a)} order parameter and {\bf (b)} its variance across bond configurations as the function of disorder probability $p$ at $T=0$ in the RBIM, adapted from \cite{LeeNishimori2022}. The order parameter $O$ is squared magnetization normalized to be within $[0,1]$. The black dashed line is the critical point at $p_c\,{=}\,0.103$~\cite{WangPreskill2003}.  }
    \label{fig:chi}
\end{figure}

In \figref{fig:chi}, the variance of the order parameter is shown along the $T=0$ line of the RBIM phase diagram~\cite{WangPreskill2003}, where $\chi$ decays with the system size in both the long-range ordered and paramagnetic phase. In this plot, the order parameter $O$ is defined as the normalized squared magnetization~\cite{LeeNishimori2022}. This numerically demonstrates that in the long-range ordered phase, the fraction of paramagnetic configurations vanishes in the thermodynamic limit. Even away from $T=0$ line, as long as the ensemble is in a long-range ordered phase, this paramagnetic fraction is expected to vanish with increasing system size. 


Therefore, when it comes to the evaluation of a disorder-averaged quantity scaling sub-extensively with system size, one can assume that the majority of bond configurations are in the same phase. This is crucial in the evaluation of the coherent information of the decohered toric code state. In the final expression of the coherent information, we encountered the following term:
\begin{align}  
    f(\Delta^x_{\bmm,\ba}) := e^{-\Delta^x_{\bmm,\ba}  } \log  \frac{ \,\sum_{\bap} e^{-\Delta^x_{\bmm,\bap}  } }{  e^{-\Delta^x_{\bmm,\ba}  }  }  > 0,
\end{align} 
where $\Delta^x_{\bmm,\ba}$ is the difference between free energies of the RBIM with bond equivalence classes $(\bmm,\ba)$ and $(\bmm,\bm{0})$ at the inverse temperature $\beta_x$.

\subsection{Coherent information}

Our goal is to understand this quantity $f(\Delta^x_{\bmm,\ba})$ in different regimes. First, assume that $\beta_x\,{>}\,\beta^c$. In this case, for the typical long-range ordered configuration at the inverse temperature $\beta_x$ along the Nishimori line, this free energy difference increase with the system size such that $|\Delta^x_{\bmm,\ba}|\,{>}\,cL$ for some non-zero constant $c$ if $\ba\,{\neq}\,\bm{0}$ and $L\,{=}\,\sqrt{N}$ is the linear size of the system. Note that $\Delta^x_{\bmm,\bm{0}}\,{=}\,0$. Therefore, we get
\begin{align}
    \ba=0:&\,\,\, f(\Delta^x_{\bmm,\ba}) \leq \log (1 + 3 e^{-cL}) < 3 e^{-cL} \nonumber \\
    \ba\neq0:&\,\,\, f(\Delta^x_{\bmm,\ba}) \leq cL e^{-cL}  + e^{-cL} \log( 1 + 3 e^{-cL})   \nonumber \\
    & \,\, \qquad \qquad < (cL + 3 e^{-cL}) e^{-cL},
\end{align}
where we used that $x e^{-x}$ is monotonically decreasing for $x\,{\in}\,[1,\infty]$. Accordingly, we obtain that
\begin{align}
      \hspace{-8pt} \sum_{\textrm{LRO }\bmm,\ba} p^x_{\bmm,\bm{0}} \cdot f(\Delta^x_{\bmm,\ba})   &<  3 e^{-cL}(cL+4) \cdot \big( \hspace{-8pt} \sum_{\textrm{LRO } \bmm,\ba} p^x_{\bmm,\bm{0}} \big) \nonumber \\
    & \xrightarrow{L \rightarrow \infty} 0,
\end{align}
since the sum of probabilities is upper-bounded by 1. 

Similarly, consider bond configurations that are paramagnetic, whose fraction $ \sum_{\textrm{para } \bmm} p^x_{\bmm,\bm{0}} \,{\sim}\,\cO(1/N)$. For these configurations, the free energy difference $\Delta^x_{\bmm,\ba}$ is upper-bounded by constant, denoted as $\Delta^x_{\bmm,\ba}\,{<}\,c'$. In this case, one can show that   
\begin{align}
    &  \sum_{\ba} f(\Delta^x_{\bmm,\ba})   \leq 2 \log 2 \quad \textrm{for para. $\bmm$}\nonumber \\
    &\Rightarrow \quad  \lim_{N \rightarrow \infty}  \hspace{-8pt} \sum_{\textrm{para }\bmm,\ba} p^x_{\bmm,\bm{0}} \cdot f(\Delta^x_{\bmm,\ba})   = 0.
\end{align}
Therefore, we establish the limiting behavior.

On the other hand, if there is a finite fraction $f\,{>}\,0$ of the bond configurations with paramagnetic order with $\Delta^x_{\bmm,\ba}\,{<}\,c'$ at any system size, one can show that 
\begin{align}
    &  \sum_{\ba} f(\Delta^x_{\bmm,\ba})  \geq 3 c' e^{-c'} \quad \textrm{for para. $\bmm$}\nonumber \\
    &\Rightarrow  \,\,\, \hspace{-8pt} \sum_{\textrm{para }\bmm,\ba} p^x_{\bmm,\bm{0}} \cdot f(\Delta^x_{\bmm,\ba})  \geq 3 f c' e^{-c'} > 0.
\end{align}
Accordingly, the coherent information will be strictly smaller than $2 \log 2$ by an $\cO(1)$ number. 

A similar argument can be made in the paramagnetic phase at $\beta_x\,{<}\,\beta^c$, where the majority of the bond configurations are paramagnetic. In fact, in the paramagnetic configuration the free energy difference is upper-bounded by a quantity $c'$ which should vanish in the thermodynamic limit, i.e.,
\begin{align}
    \beta_x < \beta_c \quad \Rightarrow \quad \lim_{N \rightarrow \infty} c' = 0.
\end{align}
Accordingly, $\lim_{N \rightarrow \infty}   \sum_{\ba} f(\Delta^x_{\bmm,\ba})   = - 2 \log 2$.

\section{Free energy as lower bound} \label{app:free_energy}

Here, we remark that the divergence of the domain wall free energy provides a lower bound on the failure probability of the maximum entropy decoder. Given syndrome observations $(\bmm,\tbm)$,  we can obtain probabilities for different values of $(\ba, \bb)$ by calculating the partition function for the associated random bond model. In the maximum entropy decoder, we correct errors based on these probabilities.

When we have an error $(\bmm,\tbm,\ba,\bb)$, the decoding is successful only if we infer the same error class. We can fail in multiple different ways; assume what we inferred is $(\bmm,\tbm,\ba',\bb')$. Let $\bmeta_1=\ba'-\ba$ and $\bmeta_2=\bb'-\bb$. The probability of failing in this way is given as
\begin{align} 
   P^{\textrm{fail}, ( \bmeta_1, \bmeta_2) }_{\bmm,\tbm, \ba, \bb} =  \frac{ P_{\bmm,\tbm,\ba+\bmeta_1,\bb+\bmeta_2} }{ \sum_{\ba',\bb'}P_{\bmm,\tbm,\ba',\bb'} },
\end{align}
where $P_{\bmm,\tbm,\ba,\bb} = p^z_{\bmm,\bm{a}} p^x_{\tbm,\bm{b}}$ is the probability of getting this error under independent $Z$ and $X$ errors. 
Now, we define $\Delta_{\bmm,\tbm,\ba,\bb}^{( \bmeta_1, \bmeta_2)}$ as the free energy of inserting a domain wall $( \bmeta_1, \bmeta_2)$ with respect to $(\bmm,\tbm,\ba,\bb)$. Then
\begin{align}
    \frac{1}{2^k}   e^{-\Delta_{\bmm,\tbm,\ba,\bb}^{( \bmeta_1, \bmeta_2)}} \leq P^{\textrm{fail}, ( \bmeta_1, \bmeta_2) }_{\bmm,\tbm, \ba, \bb} \leq 1
\end{align}
where $\Delta_{\bmm,\tbm,\ba,\bb}^{( \bmeta_1, \bmeta_2)}$ decomposes as 
\begin{align}
    \Delta_{\bmm,\tbm,\ba,\bb}^{( \bmeta_1, \bmeta_2)} = \Delta_{\bmm, \ba }^{\bmeta_1} \cdot \Delta_{ \tbm ,\bb}^{\bmeta_2} 
\end{align}
under independent $X$ and $Z$ errors. 
To further proceed, we average it over all syndrome probabilities. 
Using the Jensen's inequality, we obtain that
\begin{align} \label{eq:condition1}
   \overline{ P^{\textrm{fail}} } \geq  \frac{1}{2^k} \sum_{\bmeta_1,\bmeta_2 \neq (0,0) }  \hspace{-10pt} e^{- \langle \Delta_{\bmeta_1}^z \rangle } e^{- \langle \Delta_{\bmeta_2}^x \rangle }  .
\end{align}
Therefore, the probability of the decoding failure is lower bounded by the disorder-averaged free energy of the random bond Ising model. (If we sum over $\bmeta_i$, the prefactor $2^{-k}$ approximately gets canceled) If the free energy $\Delta$ of non-trivial domain wall insertion does not diverge with system size, the probability of decoding failure is non-zero. However, if the free energy diverges, it does not guarantee that the failure probability is non-zero since this only provides a rigorous lower bound, not an upper bound.

\section{Correlated models} \label{app:correlated}

In the main text, we showed that the density matrix of the system combined with reference qubits can be diagonalized as
\begin{align} \label{eq:DM_diagonal}
    \cE[\rho_{RQ}]  &= - \sum_{\bm{m},\ba \vphantom{\tbm,\bb}} \sum_{\tbm, \bb} 
        \, p_{\bmm \vphantom{\tilde{\bmm}}, {\ba}}^z \,p_{\tilde{\bmm},{\bb}}^x |\bmm,\ba,\tbm,\bb \rangle \langle  \bmm,\ba,\tbm,\bb |
\end{align}
by choosing an appropriate basis for logical and reference qubits as discussed in the main text. Although it is straightforward to see that Pauli-$X$ and $Z$ errors keep the density matrix diagonal in the stabilizer basis $(\bmm,\tbm)$, it is noteworthy that one can also find a basis for the \emph{logical} and \emph{reference} qubits that maintains this diagonal structure once decoherence is applied.

Moreover, the form of \eqnref{eq:DM_diagonal} remains valid even if Pauli-$X$ and $Z$ errors become correlated. In that case, instead of a product distribution $p_{\bmm \vphantom{\tilde{\bmm}}, {\ba}}^z \,p_{\tilde{\bmm},{\bb}}^x $, one obtains a correlated probability distribution $p_{\bmm \vphantom{\tilde{\bmm}}, {\ba},\tilde{\bmm},{\bb}}$ which causes an initial configuration to flip from $(\bm{0},\bm{0},\bm{0},\bm{0})$ into the basis $(\bmm \vphantom{\tilde{\bmm}}, {\ba},\tilde{\bmm},{\bb})$. For example, consider a decoherence channel
\begin{align}
    {\cal E}_i[\rho] = (1-p) \rho + \sum p_a \sigma^a_i \rho \sigma_i^a
\end{align}
where $a$ is label for Pauli $X$, $Y$, and $Z$ errors, $\sigma^a$ is Pauli matrix, and $p = \sum_a p_a$. With this notation, we obtain that
\begin{align} \label{eq:generalP}
    p_{\bmm \vphantom{\tilde{\bmm}}, {\ba},\tilde{\bmm},{\bb}} &=  (1-p)^{n} \sum_{ \bm{l} } \prod_{i=x,y,z} \qty(\frac{p_i}{1-p} )^{|\bm{l}_i|} 
\end{align}
where $\bm{l}$ is a $N$-dimensional vector, whose $i$-th entry represents which Kraus operator (1,$X$,$Y$,$Z$) is applied on the site-$i$. Thus it decomposes into $\bm{l} = \bm{l}_x + \bm{l}_y + \bm{l}_z$. Note that for each edge, only one of $\{l_{x,e}, l_{y,e}, l_{z,e}\}$ can take a non-zero value.

Given $(\bmm \vphantom{\tilde{\bmm}}, {\ba},\tilde{\bmm},{\bb})$, we can parameterized all possible $X$ and $Z$ error strings consistent with this as
\begin{align}
    \tilde{\bm{l}}_z &= \tilde{\bm{l}}_z^{\bmm,\ba} + \rd_\perp \bm{\sigma} \in \mathbb{F}_2^N \nonumber \\
    \tilde{\bm{l}}_x &= \tilde{\bm{l}}_x^{\tbm,\bb} + \rd \bm{\tau} \in \mathbb{F}_2^N  
\end{align}
where $\tilde{\bm{l}}_z^{\bmm,\ba}\,{:=}\,\bl^\bmm\,{+}\,\bm{L}^\perp_\ba$ and $\tilde{\bm{l}}_x^{\tbm,\bb}\,{:=}\,\bl^\tbm\,{+}\,\bm{L}_\bb$ are representative error strings for given syndromes and logical errors. Also, $\bsigma$ and $\bm{\tau}$ are additive $\mathbb{Z}_2=\{0,1\}$ variables residing on the vertices of dual square lattice and the original square lattice, respectively. 
Due to the \emph{mutually exclusive} behavior of error strings, we remark that
\begin{align}
    \bm{l}_x &= \tilde{\bm{l}}_x (\bm{1}- \tilde{\bm{l}}_z) \nonumber \\
    \bm{l}_z &= \tilde{\bm{l}}_z (\bm{1} - \tilde{\bm{l}}_x) \nonumber \\
    \bm{l}_y &= \tilde{\bm{l}}_x \tilde{\bm{l}}_z.
\end{align}
where $\bm{1} = (1,1,...,1) \in \mathbb{F}_2^N$ and the product is element-wise multiplication between two $n$-dimensional vectors. 
The product between $\tilde{\bm{l}}_x$ and $\tilde{\bm{l}}_z$ would generate interactions between $\bm{\sigma}$ and $\bm{\tau}$. Therefore, we obtain that the probability distribution for $p_{\bmm \vphantom{\tilde{\bmm}}, {\ba},\tilde{\bmm},{\bb}}$ is given as
\begin{align} \label{eq:SMcorr}
&  p_{\bmm \vphantom{\tilde{\bmm}}, {\ba},\tilde{\bmm},{\bb}}  \propto \sum_{\bm{\sigma}, \bm{\tau}} e^{-\sum_{a,e} H_{a,e}}   \\
    &H_{y,e} = \qty( -\frac{\beta_y}{2} + \frac{\beta_x}{2} + \frac{\beta_z}{2}  )  (-1)^{\tilde{\bl}_z^{\bmm,\ba}+ \tilde{\bl}_x^{\tbm,\bb}} \prod_{s \in \partial l}   \sigma_s \prod_{p \in \delta l}   \tau_p  \nonumber \\
    &H_{x, l} = \qty( \frac{\beta_z}{2} - \frac{\beta_x}{2} + \frac{\beta_y}{2}  )  (-1)^{\tilde{\bl}_x^{\tbm,\bb}} \prod_{p \in \delta l}   \tau_p \nonumber \\
    &H_{z, l} = \qty( \frac{\beta_x}{2} - \frac{\beta_z}{2} + \frac{\beta_y}{2}  )  (-1)^{\tilde{\bl}_z^{\bmm,\ba}} \prod_{s \in \partial l}   \sigma_s 
\end{align}
where $e^{-2 \beta_i} = \frac{p_i}{1 - p}$ and we implicitly converted an additive $\mathbb{Z}_2=\{0,1\}$ variable $\bsigma$ and $\bm{\tau}$ into multiplicative $\mathbb{Z}_2=\{1,-1\}$ variable.
The Hamiltonian for the correlated $X$ and $Z$ errors in \eqnref{eq:SMcorr} now have the form where two Ising models are coupled with each other via $H_y$ term. Accordingly, the coherent information under this correlated error is given as 
\begin{align}
    I_c = 2 \log 2 + \sum_{\bmm \vphantom{\tilde{\bmm}}, {\ba},\tilde{\bmm},{\bb}} p_{\bmm \vphantom{\tilde{\bmm}}, {\ba},\tilde{\bmm},{\bb}} \log \frac{p_{\bmm \vphantom{\tilde{\bmm}}, {\ba},\tilde{\bmm},{\bb}}}{p_{\bmm, \tbm}}.
\end{align}

\end{document}